\documentclass[a4paper,11pt]{article}
\usepackage[DIV12]{typearea}
\usepackage[latin1]{inputenc}
\usepackage[english]{babel}

\usepackage{cite}
\usepackage{amsfonts}
\usepackage{amsmath}    
\usepackage{amssymb}
\usepackage{cite}
\usepackage{mathrsfs}
\usepackage{pifont}
\usepackage{units}
\usepackage[small,loose]{subfigure}  
\usepackage{cancel}
\usepackage{ulem}
\usepackage[pdftex]{color}
\usepackage{slashed}
\usepackage{graphicx}

\usepackage[pdftex,colorlinks,pdfpagelabels]{hyperref}
\usepackage[figure,table]{hypcap} 
\hypersetup{
   bookmarksnumbered,
   pdfstartview={FitV},
   pdfpagemode={UseOutlines},
   pdfauthor={Rolf Kappl and Annika Reinert and Martin Wolfgang Winkler},
   pdftitle={},
   pdfsubject={scientific publication}
   ,citecolor={blue},
   linkcolor={blue},
   urlcolor={blue},
   filecolor={blue}
}

\newcommand{\eVdist}{\kern-0.06em}

\newcommand{\gv}{\:\text{G\eVdist V}}

\allowdisplaybreaks[1]

\begin{document}

\begin{titlepage}

\vspace*{-3.0cm}
\begin{flushright}
\end{flushright}

\begin{center}
{\Large\bf
  AMS-02 Antiprotons Reloaded
}
\\
\vspace{1cm}
\textbf{
Rolf Kappl,
Annika Reinert,
Martin Wolfgang Winkler
}
\\[5mm]
\textit{\small
Bethe Center for Theoretical Physics \& Physikalisches Institut der 
Universit\"at Bonn, \\
Nu{\ss}allee 12, 53115 Bonn, Germany
}
\end{center}

\vspace{1cm}

\begin{abstract}
The AMS-02 collaboration has released preliminary data on the antiproton fraction in cosmic rays. The surprisingly hard antiproton spectrum at high rigidity has triggered speculations about a possible primary antiproton component originating from dark matter annihilations. In this note, we employ newly available AMS-02 boron to carbon data to update the secondary antiproton flux within the standard two-zone diffusion model. The new background permits a considerably better fit to the measured antiproton fraction compared to previous estimates. This is mainly a consequence of the smaller slope of the diffusion coefficient favored by the new AMS-02 boron to carbon data.
\end{abstract}

\end{titlepage}

\vspace{2cm}

\section{Introduction}

The AMS-02 experiment at the International Space Station is performing a high precision measurement of charged cosmic ray fluxes. The spectrum of antiprotons is of particular interest as it provides a very sensitive probe of possible exotic sources in the galaxy. In this light, preliminary $\bar{p}/p$ data~\cite{Amsdayspbar} which indicate a nearly constant ratio at high rigidity have caused considerable excitement. Interpretations of the $\bar{p}/p$ data in terms of TeV mass dark matter annihilations have been suggested~\cite{Ibe:2015tma,Hamaguchi:2015wga,Lin:2015taa,Chen:2015kla}.\footnote{See also~\cite{Bringmann:2014lpa,Cirelli:2014lwa,Hooper:2014ysa} for recent dark matter studies using antiprotons.}

On the other hand, the $\bar{p}$ flux suffers from sizeable uncertainties related e.g. to the modeling of cosmic ray transport in the galaxy. It was pointed out in~\cite{Giesen:2015ufa,Evoli:2015vaa} that after fully accounting for uncertainties, no unambiguous excess is left in the AMS-02 $\bar{p}/p$ ratio. Still, viable configurations consistent with the data appeared somewhat at the edge of the uncertainty band provided in~\cite{Giesen:2015ufa}.

In this note we improve the calculation of the secondary $\bar{p}$ background\footnote{Secondary cosmic rays are generated by the scattering of primary cosmic rays on the interstellar matter.} within the two-zone diffusion model. This is achieved by employing the new AMS-02 boron to carbon (B/C) data~\cite{AmsdaysBC}. B/C being a secondary to primary ratio~\cite{Bethe:1939bt} can be used to estimate propagation parameters of the intergalactic transport equation. The AMS-02 data considerably extend the rigidity range compared to previous measurements and allow for a more robust determination of propagation parameters which -- at high energies -- are less affected by degeneracies. Our approach is more data-driven compared to previous work as we do not attempt to predict the fluxes of primary cosmic ray nuclei within the diffusion model. Merely, we directly take the observed primary fluxes as an input for the calculation of secondary cosmic ray production. We also make use of the updated $\bar{p}$ production cross sections~\cite{Kappl:2014hha} which were recently derived from scattering data of the NA49 experiment~\cite{Anticic:2009wd,Baatar:2012fua}. 

The secondary antiproton flux obtained with the new sets of propagation parameters tends to decrease less strongly with energy compared to previous estimates~\cite{Donato:2001ms,Donato:2008jk,Kappl:2014hha,Giesen:2015ufa}. This is mainly triggered by the smaller power law index of spatial diffusion $\delta$ favored by the AMS-02 B/C data. Intriguingly, the allowed range of $\delta$ includes the theoretically favored values $\delta=0.33$ and $\delta=0.5$ which correspond to a Kolmogorov and Kraichnan type power spectrum of the galactic magnetic field.

The harder antiproton spectrum following from our propagation analysis considerably improves the consistency of the AMS-02 $\bar{p}/p$ data with the background hypothesis.

\section{Transport of Cosmic Rays}

The propagation of all cosmic ray species can be performed in a unified framework. It is common to start with a parameterization of primary sources and then to calculate the fluxes of all primary and secondary species resulting from the network of scattering and spallation reactions in the galactic disc~\cite{Maurin:2001sj,Moskalenko:2001ya,Evoli:2008dv,Putze:2010zn}. As we are mainly concerned with the secondary cosmic ray species B and $\bar{p}$, we instead perform a more data-driven approach and use observed primary fluxes as an input for our calculation. Before discussing this in more detail, we will very briefly review the relevant aspects of cosmic ray propagation.

\subsection{Propagation in the Diffusion Model}

Cosmic rays propagating through the galaxy are affected by various processes which are encoded in the diffusion equation. The latter relates the production rate (source term) $q_i$ of the cosmic ray species $i$ to the resulting space-energy density $N_i$
\begin{equation}
\label{eq:diffusionequation}
 \nabla (-K \:\nabla N_i + \boldsymbol{V}_c \,N_i) + 
\partial_E (b_\text{tot} \,N_i -K_{EE} \:\partial_E N_i ) 
+ \Gamma_\text{ann}\,N_i = q_i\;, 
\end{equation} 
where $E$ stands for the total energy. $K$ accounts for diffusion on magnetic field inhomogeneities. The galactic wind \(\boldsymbol{V}_c\) is responsible for convection. The function \(b_\text{tot}\) includes Coulomb, ionization and adiabatic energy losses as well as energy losses induced by reacceleration. \(K_{EE}\) is the energy diffusion coefficient and \(\Gamma_\text{ann}\) the annihilation rate on the interstellar matter~\cite{Strong:1998pw,Maurin:2002ua}.

In this work we employ the two-zone diffusion model~\cite{Maurin:2001sj,Donato:2001ms} in which diffusion occurs homogeneously in a cylinder of half-height $L$ around the galactic disc.\footnote{See~\cite{Strong:1998pw,Moskalenko:2001ya,Evoli:2008dv} for alternative approaches to solve the diffusion equation.} \(K\) is parameterized as
\begin{equation}
 K=K_0\beta\left(\frac{\mathcal{R}}{\text{GV}}\right)^\delta\;,
\end{equation}
where $K_0$ and $\delta$ are the normalization and the power law index of spatial diffusion. Magnetohydrodynamics considerations suggest values $\delta=0.33$ (Kolmogorov spectrum) or $\delta=0.5$ (Kraichnan spectrum) depending on the detailed modeling of magnetic field turbulences (see e.g.~\cite{Strong:2007nh}).
The velocity and rigidity of the cosmic ray particle are denoted by $\beta$ and $\mathcal{R}$. The convective wind \(\boldsymbol{V}_c\) is taken to be constant in the diffusion cylinder and pointing away from the galactic disc. Finally, the Alfv\'en speed of magnetic shock waves $V_a$ enters $K_{EE}$ and completes the set of five independent propagation parameters: $L$, $K_0$, $\delta$, $V_c$ and $V_a$. 

\subsection{Source Terms}
A common approach to predict cosmic ray spectra is to start from a parameterization of primary sources. A standard choice is $q_i^\text{prim}\propto \beta^\eta \,\mathcal{R}^\gamma$ with $\eta=0$~\cite{Maurin:2001sj} or $\eta=1$~\cite{Putze:2010zn} (see also~\cite{Putze:2010fr}). At high energies, this leads to a power law form of the primary flux $\Phi_i \propto \mathcal{R}^{\gamma+\delta}$. The index $\gamma$ is matched to observed high energy fluxes, e.g.\ $\gamma=2.65 -\delta$ in~\cite{Putze:2010zn}. Primary fluxes over the full energy range are then calculated by means of the diffusion equation and seeded into the source terms for secondaries. 

One drawback of this approach is that the low energy parameterization of primary sources is rather ambiguous. The observed spectral breaks in the proton and helium spectra~\cite{Amsdayshelium,Aguilar:2015ooa} further indicate that a simple power law may not be sufficient to model the primary sources at high energies. For this reason, we have chosen not to start from a parameterization of primary sources, but directly use the observed primary fluxes as an input for the calculation of secondaries.

Secondary cosmic rays like B or $\bar{p}$ are generated by scattering of (mainly) primary cosmic rays on the interstellar matter. The source term for secondary production reads
\begin{equation}
\label{eq:source}
 q_i^\text{sec}(T) = \sum\limits_{j}\;\,4 \pi\int dT' 
\left[n_\text{H}\left(\frac{d\sigma}{dT}\right)_{j\text{H}\rightarrow i}
+ n_\text{He}\left(\frac{d\sigma}{dT}\right)_{j\text{He}\rightarrow i}
\right] \;\;\Phi_j^\text{IS}(T')\;.
\end{equation}
Here $\Phi_j^\text{IS}(T')$ is the interstellar flux of the incoming cosmic ray species $j$ as a function of its kinetic energy per nucleon $T'$, while $T$ denotes the kinetic energy per nucleon of the outgoing secondary particle $i$. The differential cross section $d\sigma/dT$ for the production of $i$ includes the prompt process as well as the production via intermediate unstable particles which decay into $i$. We consider only scattering processes off hydrogen and helium in the galactic disc and use constant values $n_{\text{H}}=0.9\:\text{cm}^{-3}$ and $n_{\text{He}}=0.1\:\text{cm}^{-3}$ for their number densities~\cite{Putze:2010zn}. Furthermore, we take $q_i^\text{sec}$ to be radially constant in the disc. Local secondary cosmic ray fluxes are not particularly sensitive to the last two assumptions as long as they are used in the same way for the determination of propagation parameters. 

\subsection{Fluxes of Primary Cosmic Rays}

The interstellar fluxes of primary cosmic rays 
have been measured over a wide energy range by various experiments~\cite{Maurin:2013lwa}. 
As experimental data refer to the top-of-the-atmosphere (TOA) fluxes we have to take into account solar modulation. We ignore charge-sign dependent effects and use the force field approximation~\cite{Gleeson:1968zza}. The value of the force field $\phi$ for experimental data prior to 2011 is extracted from~\cite{2011JGRA..116.2104U}. The extended table~\cite{usoskintable} further allows us to deduce $\phi=0.57\gv$ for the new AMS-02 data. 

\begin{figure}[ht]
\centering
\includegraphics[width=12cm]{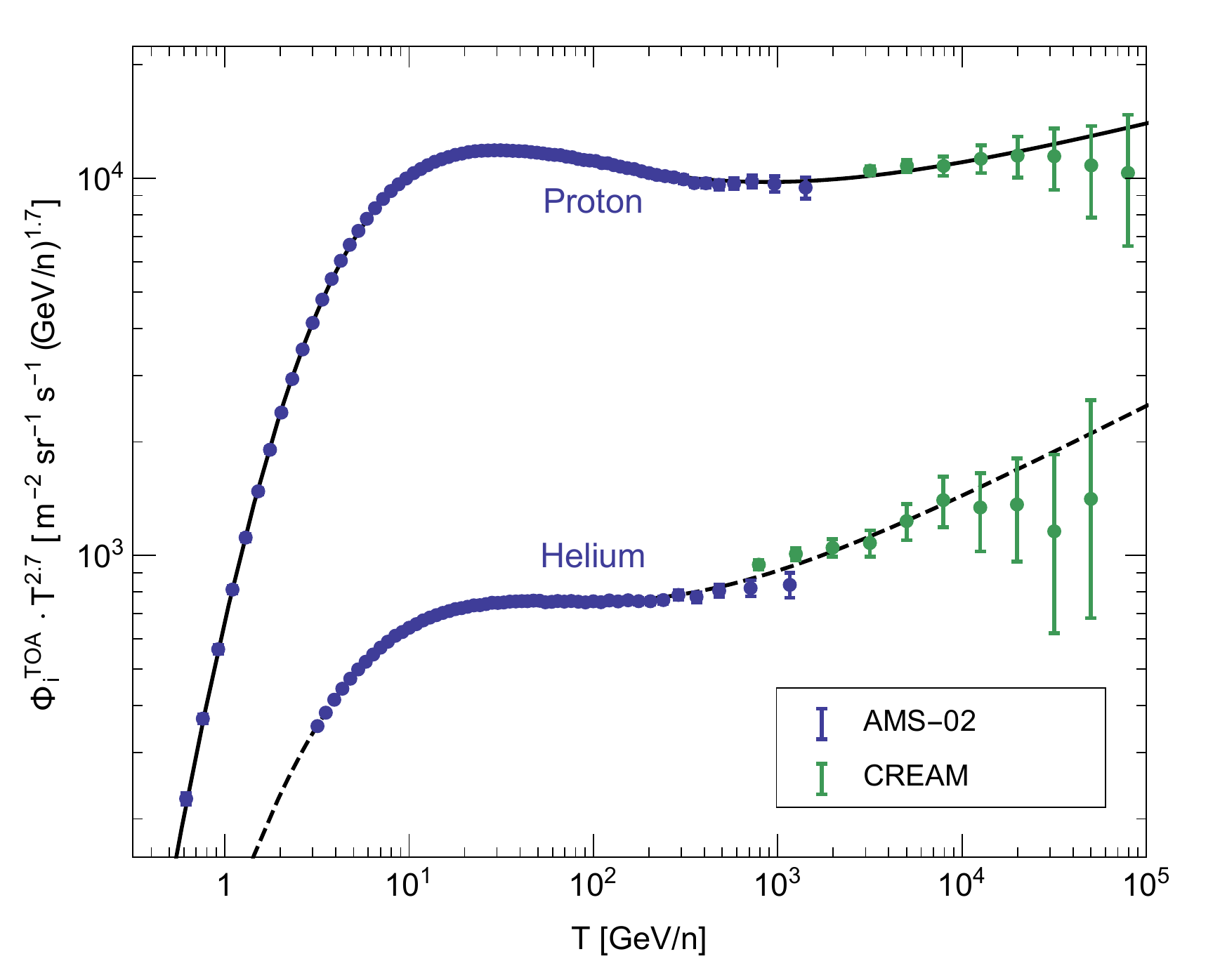}
\caption{Proton and helium fluxes measured by AMS-02~\cite{Amsdayshelium,Aguilar:2015ooa} and CREAM~\cite{Yoon:2011aa}. Also shown are the fit functions employed in this work. The latter were modulated with a force field $\phi=0.57\:\text{GV}$.}
\label{fig:protonhelium}
\end{figure}

The proton and helium fluxes released by the AMS-02 collaboration~\cite{Amsdayshelium,Aguilar:2015ooa} clearly indicate a spectral hardening at rigidity $\mathcal{R}\sim 400\:\text{GV}$ consistent with previous measurements by PAMELA~\cite{Adriani:2011cu} and CREAM~\cite{Yoon:2011aa} (see also~\cite{Vladimirov:2011rn,Tomassetti:2012ga,Thoudam:2014sta,Bernard:2012pia}). We thus parameterize the interstellar proton and helium fluxes as the sum of two power laws
\begin{align}\label{eq:HHefit}
\Phi_i^\text{IS}(T)=\left(\frac{T}{T+b}\right)^c\left(A_1T^{-\gamma_1}+A_2T^{-\gamma_2}\right).
\end{align}
The bracket which multiplies the two terms accounts for the low energy ($T\lesssim 10\:\text{GeV/n}$) behavior of the observed fluxes. We determine the free parameters by a $\chi^2$ fit to the data of AMS-02 and CREAM. The best fit values are presented in table~\ref{tab:HHefit}. The experimental data are shown with our fit function in figure~\ref{fig:protonhelium}.
\begin{table}[ht]
\centering
\begin{tabular}{lcccccc}
&\(A_1\)&\(A_2\)&\(b\)&\(c\) &\(\gamma_1\)&\(\gamma_2\)\\
\hline
\text{Protons}&53488 &3294 &2.40 &2.588 &3.168 &2.576\\
\text{Helium}&3194&141.5 &5.18 &1.482 &3.163 &2.451\\
\end{tabular}
\caption{Fit parameters for the primary spectra of protons and helium~\eqref{eq:HHefit} with $A_1, A_2$ in $\text{m}^{-2}\text{s}^{-1}\text{sr}^{-1}\text{(GeV/n)}^{-1}$ and $b$ in GeV/n.}
\label{tab:HHefit}
\end{table}

\begin{figure}[t!!]
\centering
\includegraphics[width=9.9cm]{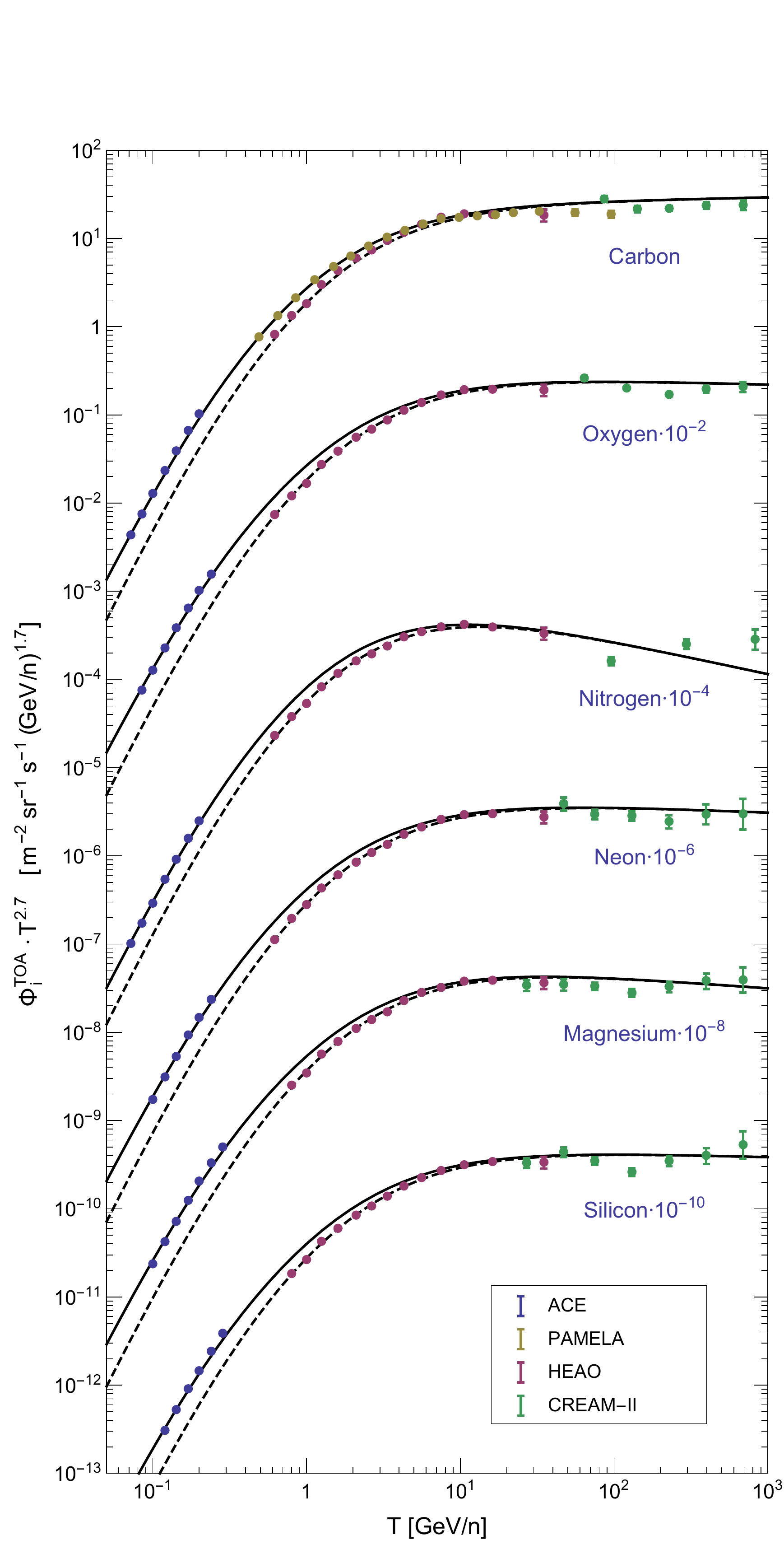}
\caption{Measured primary cosmic ray fluxes. Also shown are the fit functions employed in this work. The latter were modulated with the force field $\phi=0.43\:\text{GV}$ for ACE (solid line) and the force field $\phi=0.74\:\text{GV}$ for HEAO (dashed line).}
\label{fig:primaries}
\end{figure}

For the secondary production of boron the most relevant process is the spallation of carbon in the interstellar medium. However, several other elements, namely oxygen, nitrogen, neon, magnesium and silicon contribute significantly~\cite{Maurin:2002ua}.
The primary fluxes are determined from a fit to the experimental data below $1000\:\text{GeV/n}$. In order to cover the low, intermediate and high energy range, we extract the measurements of ACE~\cite{ace1}, HEAO~\cite{heao} and CREAM-II~\cite{Ahn:2009tb} from the cosmic ray database~\cite{Maurin:2013lwa}. For the carbon flux an additional data set from PAMELA~\cite{Adriani:2014xoa} is considered. Primaries above $1000\:\text{GeV/n}$ are excluded from our fit as they do not contribute to the boron flux in the energy range accessible to AMS-02. The interstellar flux as a function of the kinetic energy per nucleon $T$ is modeled as in the case of protons and helium, but without the second power law (as the presence of a spectral break is not yet statistically significant)
\begin{align}\label{eq:Bprimariesfit}
\Phi^{\text{IS}}_{i}(T)=A\left(\frac{T}{T+b}\right)^c T^{-\gamma}.
\end{align}
In order to avoid a too strong impact of low energy data points, suffering from uncertainties like solar modulation, we first determine the power law index from the data above $10\:\text{GeV/n}$. In the second step, we fit the remaining parameters to the full data set. The best fit parameters for oxygen, nitrogen, carbon, neon, magnesium and silicon are presented in table \ref{tab:Bprimaries} and the corresponding fluxes are displayed in figure \ref{fig:primaries}.

\begin{table}[ht]
\centering
\begin{tabular}{lcccc}
&\(A\)&\(b\)&\(c\) &\(\gamma\)\\
\hline
\text{Carbon}&23.13&1.12&2.086&2.666\\
\text{Oxygen}&30.90&1.76&1.844&2.748\\
\text{Nitrogen}&15.67&1.39&2.759&3.077\\
\text{Neon}&5.01&1.57&2.035&2.769\\
\text{Magnesium}&7.72&1.97&1.927&2.829\\
\text{Silicon}&5.33&2.04&1.822&2.746\\
\end{tabular}
\caption{Fit parameters for the primary spectra described by \eqref{eq:Bprimariesfit} with $A$ in $\text{m}^{-2}\text{s}^{-1}\text{sr}^{-1}\text{(GeV/n)}^{-1}$ and $b$ in GeV/n.}
\label{tab:Bprimaries}
\end{table}

\section{Propagation Parameters from a B/C Analysis}

In the next step, we wish to determine sets of propagation parameters consistent with the preliminary B/C spectrum released by AMS-02~\cite{AmsdaysBC}. For the calculation of the $^{11}$B and $^{10}$B source terms we consider spallation of the cosmic ray species listed in table~\ref{tab:Bprimaries} which account for $\sim98\%$ of the boron production~\cite{Maurin:2002ua}. The isotopic composition of the incoming cosmic rays is extracted from the cosmic ray data base~\cite{Maurin:2013lwa}.

For the nuclear spallation cross sections, we use the straight-ahead approximation which implies that the kinetic energy per nucleon is preserved during a spallation process. Cross sections of nuclei on hydrogen are taken from~\cite{Webber03} (as provided in the DRAGON code~\cite{Evoli:2008dv}), the energy-dependent enhancement factor for spallation on helium is extracted from~\cite{Ferrando:1988tw}. We include the indirect production of boron via intermediate radioactive isotopes in the spallation cross sections and also account for the production of $^{10}$B via neutron stripping of $^{11}$B. For the annihilation cross sections on the interstellar hydrogen and helium we use~\cite{Tripathi1999349,Tripathi:1999nw}.

Spallation cross sections suffer from uncertainties related to the lack of experimental data at high collision energies (see~\cite{Genolini:2015cta} for a comprehensive discussion). In order to model this uncertainty, we follow~\cite{Maurin:2010zp} and allow for a systematic energy bias in the parameterization of spallation cross sections. This is done by multiplying cross sections by a factor $[T/(\text{GeV/n})]^\Delta$ at $T>1\:\text{GeV/n}$ which goes into a constant at $T=10\:\text{GeV/n}$. As can be seen in figure~3 of~\cite{Maurin:2010zp} any choice between $\Delta=-0.05$ and $\Delta=0.05$ is consistent with existing experimental data.

For our analysis, we generate a random set of propagation parameters within the intervals $L=2-15\:\text{kpc}$, $\delta=0.2-0.9$, $V_c=0-20\:\text{km}\ \text{s}^{-1}$, $V_a=0-100\:\text{km}\ \text{s}^{-1}$ and free $K_0$. For the given configuration we determine the B/C ratio after accounting for solar modulation. Then, we perform a $\chi^2$ test against the B/C data measured by AMS-02~\cite{AmsdaysBC}. The total experimental uncertainty is extracted from slide 14 of~\cite{AmsdaysBC} by adding errors in quadrature. In order to account for the uncertainty in the spallation cross section, we allow the configuration to choose any value between $\Delta=-0.05$ and $\Delta=0.05$. A configuration is selected if $\chi^2/\text{d.o.f.}<2$ for the optimal $\Delta$.

\begin{figure}[t]
\centering
\includegraphics[width=14cm]{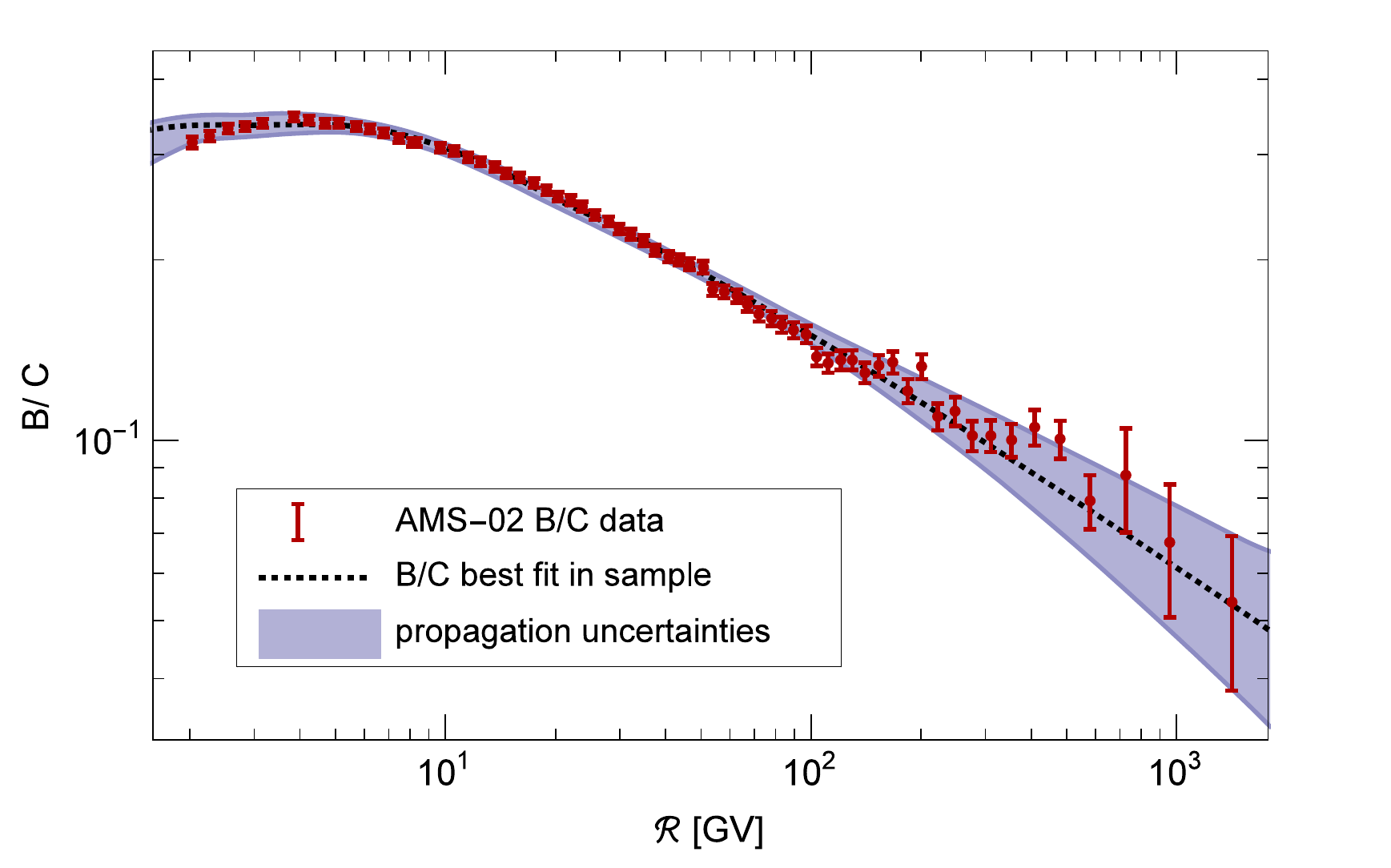}
\caption{Envelope of B/C ratios within the sample of 500 selected configurations (see text) together with the experimental data of AMS-02. The B/C ratio for the best fit configuration in the sample (table~\ref{tab:bestBC}) is indicated by the dotted line.}
\label{fig:BCscan}
\end{figure}

In figure~\ref{fig:BCscan} we depict the envelope of B/C ratios within a sample of 500 selected configurations. The set of propagation parameters which yields the best fit among this sample ($\chi^2/\text{d.o.f.}=1.2$) is given in table~\ref{tab:bestBC}.

\begin{table}[htb]
\centering
\begin{tabular}{ccccc}
\(\delta\)&\(K_0\ (\text{kpc}^2\ \text{Myr}^{-1})\)
&\(L\ (\text{kpc})\)&\(V_c\ (\text{km}\ \text{s}^{-1})\)&\(V_a\ 
(\text{km}\ \text{s}^{-1})\)\\
\hline
0.408&0.0967&13.7&0.2&31.9
\end{tabular}
\caption{Propagation parameters giving the best fit to the AMS-02 B/C data among a sample of 500 selected configurations (see text).}
\label{tab:bestBC}
\end{table}

As the presence of convective winds in the galaxy is speculative~\cite{Strong:2007nh} it is remarkable that the best configuration has almost vanishing convective wind. This already implies that the AMS-02 B/C data are consistent with a pure diffusion-reacceleration model of cosmic ray propagation. The value of $\delta$ is of particular interest as it sets the slope of secondary to primary ratios at high rigidity. In figure~\ref{fig:scatter} we show $\delta$ and $V_c$ for our sample of configurations. There is an obvious correlation between the two parameters: larger values of $\delta$ require a higher convective wind. Despite the scatter and the difference in the statistical method, we can make out a clear trend if we compare our results with previous B/C analyses in the two-zone diffusion model~\cite{Maurin:2001sj,Putze:2010zn}. The allowed parameter range lies at smaller $\delta$. In our sample we find a median of $\delta=0.48$, while the best fit configuration of~\cite{Putze:2010zn} has $\delta=0.86$. The mentioned analysis is, however, based on earlier B/C data sets. We have checked that the best fit configurations of~\cite{Maurin:2001sj,Putze:2010zn} substantially underestimate the AMS-02 B/C ratio at high rigidity. As cosmic ray transport at high energies is dominated by diffusion, the B/C ratio in this regime is particularly important for a robust determination of $\delta$. The good accuracy of the AMS-02 data at high rigidity gives us confidence that the shift towards lower $\delta$ will be confirmed by future data sets. This is further substantiated as analyses of the PAMELA~\cite{Adriani:2014xoa} and earlier AMS-02 B/C data sets~\cite{Genolini:2015cta} which were performed in slightly different propagation models yield a range of $\delta$ compatible with our findings.

The range of $\delta$ found in our analysis includes the theoretically favored values $\delta=0.33$ and $\delta=0.5$. We will show in the next section that the lower $\delta$ does also have important implications for the antiproton fraction.

\section{Antiproton Fraction}

Secondary antiprotons originate from the scattering of protons and helium on the interstellar matter. The primary fluxes are given in~\eqref{eq:HHefit} and the fit parameters in table~\ref{tab:HHefit}. Contributions from heavier nuclei are negligible. 

We make use of the new calculation of antiproton production cross sections performed in~\cite{Kappl:2014hha} (for other recent approaches see~\cite{diMauro:2014zea,Kachelriess:2015wpa}). Compared to previous parameterizations~\cite{Tan:1982nc,Duperray:2003bd} it contains a detailed treatment of antiproton production via hyperon decay as well as possible isospin effects in antineutron production. Further the description of processes involving helium has been improved. We follow~\cite{Kappl:2011jw} and include tertiary antiproton production by inelastic scattering of secondary antiprotons on the interstellar matter with annihilation and inelastic cross sections taken from~\cite{Protheroe:1981gj,Tan:1983de}.

\begin{figure}[t]
\centering
\includegraphics[width=14cm]{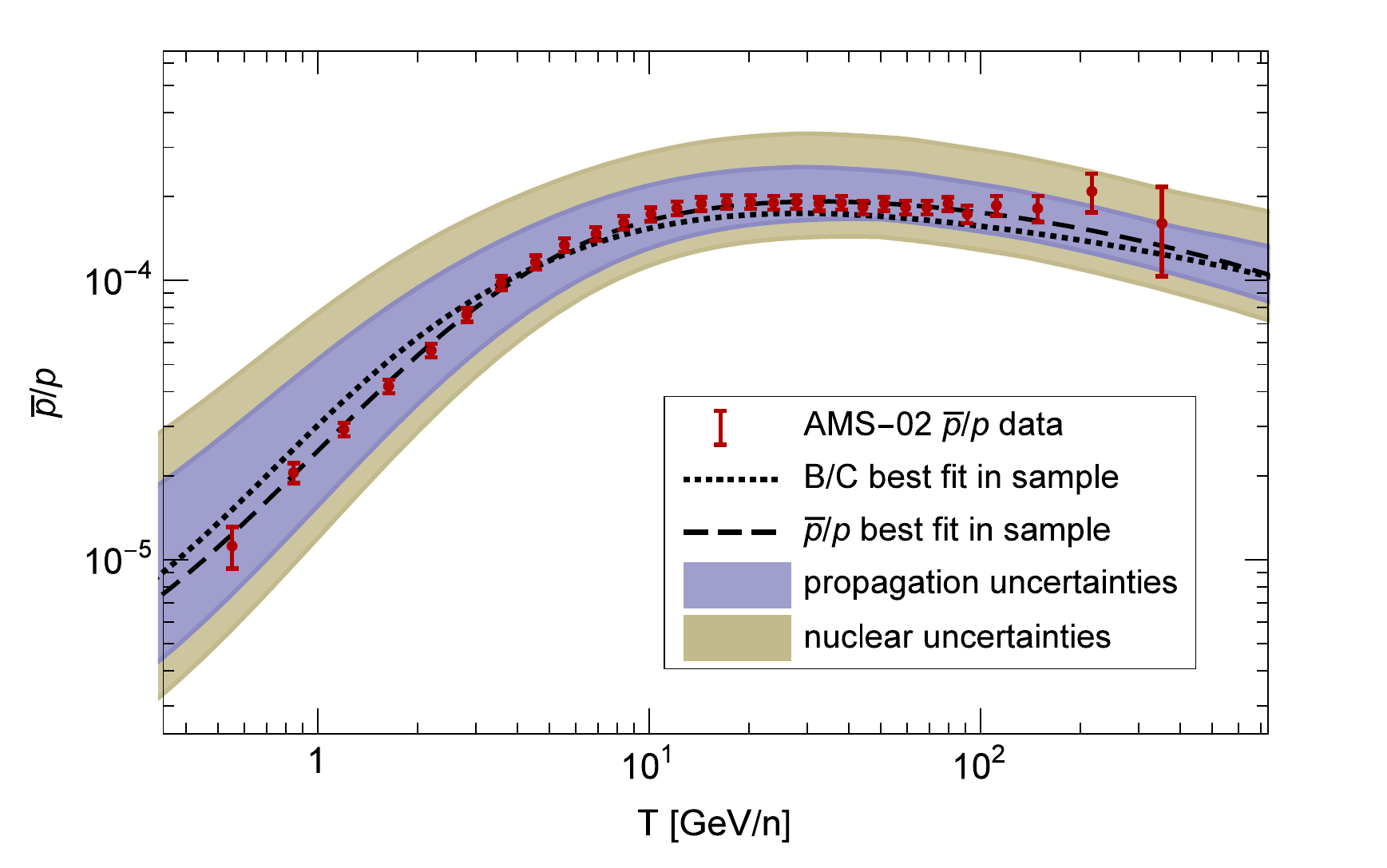}
\caption{Antiproton fraction predicted from pure secondary production compared to the AMS-02 data. The inner band encompasses propagation uncertainties (see text), the full band also includes uncertainties in the $\bar{p}$ production cross sections. The antiproton fraction for the propagation configuration (within our sample) which yields the best fit to B/C  (table~\ref{tab:bestBC}) and for the configuration which yields the best fit to the $\bar{p}/p$ data are indicated by the dotted and the dashed line, respectively.}
\label{fig:pbarp}
\end{figure}

For the sample of 500 configurations consistent with B/C (see previous section) we determine the interstellar $\bar{p}$ flux. We take the envelope of all resulting fluxes to model the propagation uncertainties. The corresponding $\bar{p}/p$ ratio after accounting for solar modulation ($\phi=0.57\:\text{GV}$) is shown with the AMS-02 data in figure~\ref{fig:pbarp}. The broader band in the same figure is obtained by including the uncertainties in the antiproton production cross sections from~\cite{Kappl:2014hha}. The $\bar{p}/p$ ratio for the configuration of table~\ref{tab:bestBC} is also shown.

It can be seen that the secondary antiproton background is in good agreement with the data, primary sources of antiprotons are not favored. To make this more explicit, we have performed a $\chi^2$ test against the AMS-02 $\bar{p}/p$ data for each configuration within our sample. Even before taking into account the uncertainties in the antiproton production cross section, we find a configuration with $\chi^2/\text{d.o.f.}$ as low as 0.5. The $\bar{p}/p$ ratio for this configuration is also shown in figure~\ref{fig:pbarp}.

\begin{figure}[tb]
\centering
\includegraphics[width=12cm]{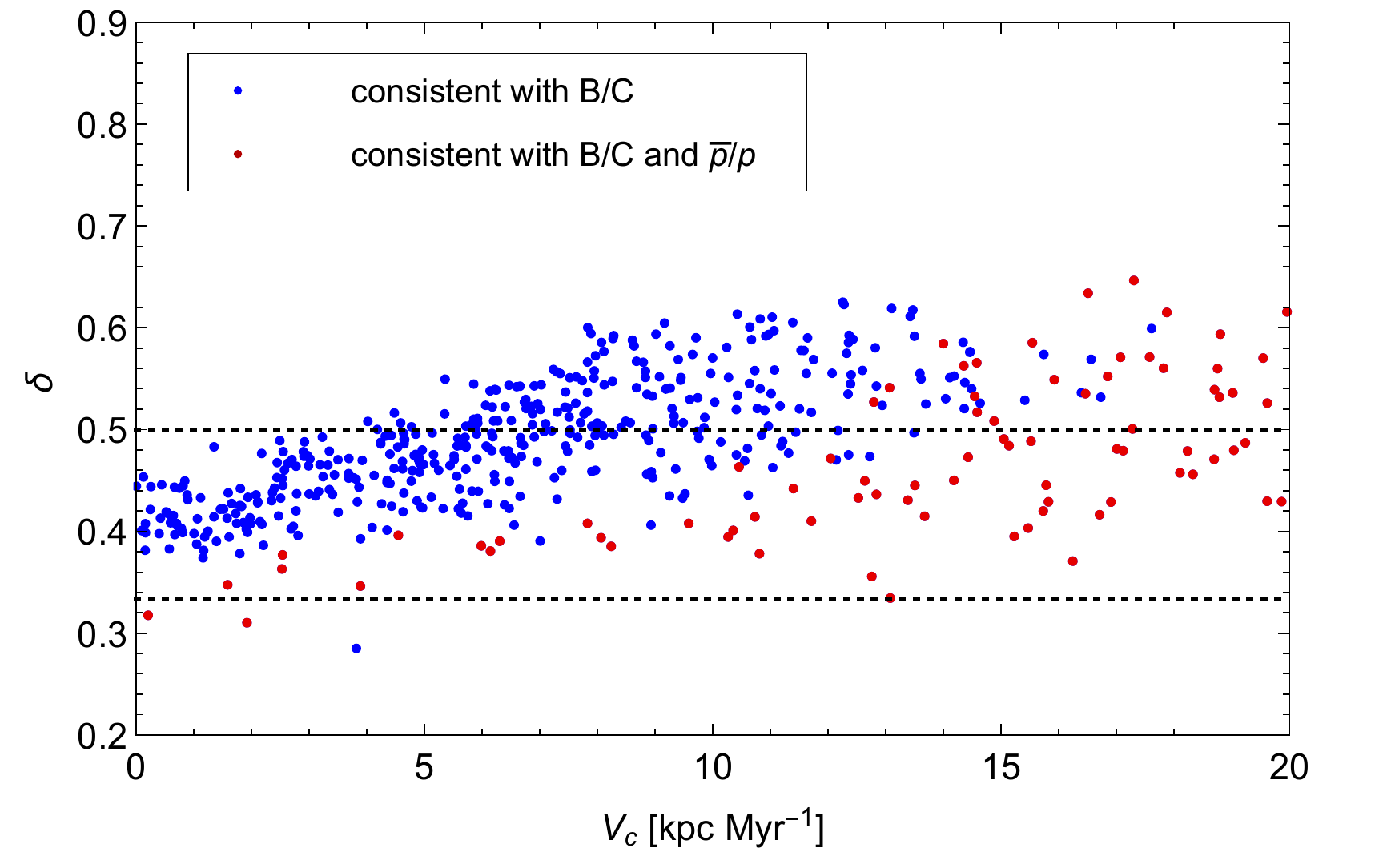}
\caption{Colored points indicate values of $\delta$ and $V_c$ within a sample of 500 configurations consistent with the AMS-02 B/C data. Configurations which simultaneously fit the AMS-02 $\bar{p}/p$ data are displayed in red. Theoretically preferred values of $\delta=0.33$ and $\delta=0.5$ are indicated by the dashed lines.}
\label{fig:scatter}
\end{figure}

In figure~\ref{fig:scatter}, which shows $\delta$ and $V_c$ for the sample of configurations selected in the B/C analysis, we have marked those which are also consistent with the $\bar{p}/p$ data. As a criterion we again required\footnote{We also took into account the uncertainty in the production cross section. For each set of propagation parameters, we calculated the minimal, medium and maximal flux within the cross section uncertainty band. Then, we defined a parameter which smoothly interpolates between the three fluxes and selected the parameter which minimizes $\chi^2$. If for this optimal choice of the production cross section $\chi^2/\text{d.o.f.}<2$ the configuration is taken to be consistent with the $\bar{p}/p$ data of AMS-02.} $\chi^2/\text{d.o.f.}<2$.

There is a trend that the AMS-02 $\bar{p}/p$ data favor those sets of propagation parameters with smaller $\delta$. This is a consequence of the almost flat shape of the AMS-02 $\bar{p}/p$ ratio at high rigidity. Both AMS-02 data sets, B/C and $\bar{p}/p$, are consistent with Kolmogorov ($\delta=0.33$) and with Kraichnan ($\delta=0.5$) type diffusion. In the case of $\delta=0.33$ convective winds are not required.

While we have shown that the AMS-02 $\bar{p}/p$ data are consistent with the background hypothesis, the relatively large uncertainties still leave considerable room for a possible primary $\bar{p}$ component. In order to improve the search for dark matter in cosmic rays, it will be crucial to shrink the error bands in the background. Among other things, this requires new experimental data on nuclear spallation cross sections which are a limiting factor in determining the propagation parameters.

\section{Conclusion}

We investigated whether the preliminary AMS-02 $\bar{p}/p$ data can be explained by the secondary antiproton background within the two-zone diffusion model of cosmic ray propagation. In the first step we determined viable sets of propagation parameters by use of the new AMS-02 B/C data. Compared to previous analyses, we found a trend towards smaller slopes of the diffusion index. The theoretically favored values $\delta=0.33$ and $\delta=0.5$ are both consistent with the data.

We then determined the error band of the $\bar{p}/p$ ratio by including uncertainties in the propagation parameters as well as in the cross sections for antiproton production. The AMS-02 $\bar{p}/p$ data reside well within the error band. No primary sources of antiprotons like dark matter annihilations are required. Indeed, the background we obtained gives a somewhat better fit to the AMS-02 data compared to the background of~\cite{Giesen:2015ufa}. This is mainly a consequence of the smaller $\delta$ favored by our propagation analysis, which translates into a harder antiproton spectrum at high rigidity. 

AMS-02 is expected to further extend the energy range and the accuracy of the antiproton data within the next years. The secondary antiproton ratio and the uncertainty band provided in this work will allow an improved background modeling in future dark matter studies.

Upcoming AMS-02 data on primary and secondary fluxes will further reduce the uncertainties in cosmic ray propagation. In the near future, uncertainties in the cross sections entering secondary source terms will be a limiting factor. This includes nuclear spallation cross sections which are not well-measured at high energies as well as the antiproton production cross sections (in particular the component from antineutron  decay). Future dark matter searches in cosmic rays will aim at identifying a sub-dominant signal on top of a large astrophysical background. In this light, there is an urgent need for a dedicated experimental program~\cite{need} to measure cross sections in the relevant energy range.

\section*{Acknowledgments}
This work has been supported by the German Science
Foundation (DFG) within the SFB-Transregio TR33 ``The Dark Universe''.

\bibliography{pbar}
\bibliographystyle{ArXiv}
\end{document}